\documentclass{elsart4-1}

\usepackage{graphicx}
\usepackage{epsfig}

\usepackage{floatrow}

\usepackage{amssymb}

\usepackage[english,francais]{babel}

\usepackage{url}

\newtheorem{e-proposition}[theorem]{Proposition}

\newtheorem{e-definition}[theorem]{Definition\rm}

\renewcommand{\theequation}{\arabic{equation}}
\def\og{\leavevmode\raise.3ex\hbox{$\scriptscriptstyle\langle\!\langle$~}}
\def\fg{\leavevmode\raise.3ex\hbox{~$\!\scriptscriptstyle\,\rangle\!\rangle$}}

\begin{document}

\centerline{Astrophysics}
\begin{frontmatter}



\selectlanguage{english}
\title{Space Detectors for Gamma Rays (100 MeV $-$ 100 GeV): from EGRET to Fermi LAT}


\selectlanguage{english}
\author{David J. Thompson}
\ead{David.J.Thompson@nasa.gov}

\address{Astroparticle Physics Laboratory, NASA Goddard Space Flight Center\\Greenbelt, Maryland USA}

\medskip
\begin{center}
{\small Received *****; accepted after revision +++++}
\end{center}

\begin{abstract}
The design of spaceborne high-energy (E$>$100 MeV) $\gamma$-ray detectors depends on two principal factors: (1) the basic physics of detecting and measuring the properties of the $\gamma$ rays; and (2) the constraints of operating such a detector in space for an extended period.  Improvements in technology have enabled major advances in detector performance, as illustrated by two successful instruments, EGRET on the Compton Gamma Ray Observatory and LAT on the Fermi Gamma-ray Space Telescope. 

\vskip 0.5\baselineskip

\selectlanguage{francais}
\noindent{\bf R\'esum\'e}
\vskip 0.5\baselineskip
\noindent
{\bf D\'etecteurs spatiaux de rayons gamma (100 MeV $-$ 100 GeV): depuis EGRET jusqu'\`a Fermi-LAT }

L'architecture des d\'etecteurs spatiaux de rayons gamma de hautes \'energies (E$>$100 MeV) d\'epend de deux facteurs principaux: (1) les principes physiques de base permettant la d\'etection et la mesure des propri\'et\'es des rayons gamma, et (2) les contraintes de fonctionnement de tels d\'etecteurs dans l'espace sur de longues dur\'ees. Les progr\`es techniques ont permis des avanc\'ees majeures concernant les performances des d\'etecteurs telles qu'illustr\'ees par les deux instruments que sont EGRET \`a bord du Compton Gamma Ray Observatory et LAT \`a bord du Fermi Gamma-ray Space Telescope.

\vskip 0.5\baselineskip
\noindent{\small{\it Keywords~:} Gamma rays; Detectors; Space \vskip 0.5\baselineskip
\noindent{\small{\it Mots-cl\'es~:} Rayons gamma~; D\'etecteurs~; Spatiaux}}

\end{abstract}
\end{frontmatter}


\selectlanguage{english}

\section{Introduction}
\label{sec:intro}

Earth's atmosphere is opaque to many types of electromagnetic radiation, including $\gamma$ rays.  Direct detection of these energetic photons therefore requires instruments in space.  Although $\gamma$ rays represent a broad swath of the electromagnetic spectrum, essentially everything above about 0.1 MeV, this article concentrates on the band from $\sim$100 MeV to $\sim$100 GeV, where satellite detectors have been particularly successful in revealing high-energy, nonthermal phenomena in the Universe. 

The material is organized in three parts: (1) the physics of pair production and multiple Coulomb scattering, which are central to these detectors; (2) the challenges of designing a $\gamma$-ray detector within the constraints of launching and operating an instrument in space; and (3) the  implementation of the design principles using technologies of two eras, represented by the Energetic Gamma Ray Experiment Telescope (EGRET) on the Compton Gamma Ray Observatory (CGRO) and the Large Area Telescope (LAT) on the Fermi Gamma-ray Space Telescope (Fermi). 

\section{Basic Principles}

The goal of any astrophysical instrument is to measure the properties of the incoming radiation: arrival direction, arrival time, energy, and possibly polarization.   Photons with energy E $>$ 100 MeV interact almost exclusively through the process of electron-positron pair production: $\gamma$ $\rightarrow$  $e^+$ +  $e^-$.  The photon itself ceases to  exist; therefore reflection (mirrors) and refraction (lenses) are not applicable to $\gamma$-ray detectors.  The properties of the incident $\gamma$ ray can only be derived from measurements of the two charged particles. This basic principle explains why all high-energy $\gamma$-ray detectors use instrumentation developed for nuclear and particle physics.  

Pair production cannot occur in free space; that would violate conservation of energy and momentum.  The most common agent to enable pair production is an atomic nucleus, which absorbs some of the momentum, but almost none of the energy.  The combined energy of the electron and positron is therefore a good approximation to the total energy of the original $\gamma$ ray.  

\begin{figure}[!t]
 \centering
 \includegraphics[width=7.5cm]{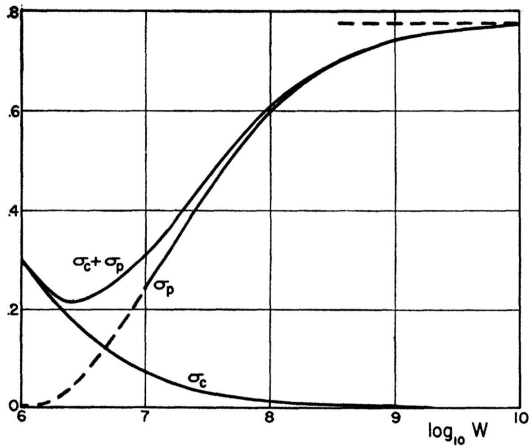}
 \caption{Probability per radiation length of lead for photons with energy W in eV to interact: $\sigma_{\mathrm{c}}$ is the probability for Compton scattering; $\sigma_{\mathrm{p}}$ is the probability for pair production. Reprinted figure with
permission from \cite{Rossi_Greisen}, copyright 1941
by the American Physical Society.}
 \label{fig:Pair_production}
\end{figure}

The first requirement for a $\gamma$-ray detector is some material to act as a converter.   
Figure~\ref{fig:Pair_production} shows the probability for photon interactions in lead as a function of energy.  For one radiation length of material, a large fraction of the high-energy photons can be converted to pairs.  Because higher-Z materials have shorter radiation lengths (in g cm$^{-2}$), choosing such materials minimizes the mass of material needed to achieve a reasonable probability of having the $\gamma$ rays detected. See \cite{Rossi},\cite{Olive} for a detailed discussion of pair production and related electromagnetic processes. 

Converting the $\gamma$ rays into electron-positron pairs is the easy part of building a detector.  Extracting the properties of the incident photons from the particle pair is harder.  Even so, two pieces of information are fairly straightforward to determine:  (1)  the arrival time of the photon can be measured by any detector, such as a scintillator, that is hit by either particle; and (2) the energy of the photon can be approximated by measuring the energies of the two particles in some sort of energy calorimeter, correcting for energy losses in any material they pass through. 

The bigger challenge is determining the arrival direction of the incident $\gamma$ ray, an essential bit of information for astrophysics.  Because the $e^+$ and  $e^-$ carry most of the energy of the photon (at least for energies well above the combined rest mass energies of the two particles), their paths after the point of conversion can be used to derive this direction.  The most probable opening angle, in radians, between the electron and positron can be approximated as \cite{Olsen}

\begin{equation}
\theta_{\mathrm{open}}=\frac{\mathrm{0.8}}{\mathrm{E_{\gamma}}}
\end{equation}

\noindent where $E_{\gamma}$ is the $\gamma$-ray energy in MeV. Above 100 MeV, this opening angle is relatively small (and decreases with increasing energy), allowing the track directions to be used to measure the photon arrival direction. The immediate requirement is therefore some sort of particle tracking detector.  

\begin{figure}[!t]
 \centering
 \includegraphics[width=10.5cm]{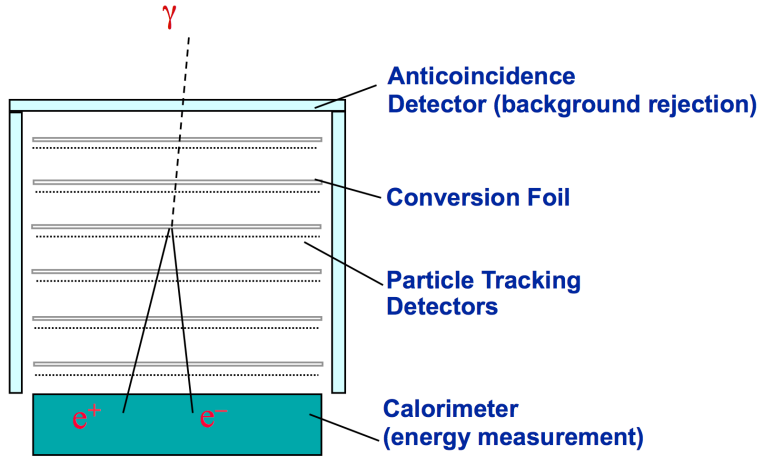}
 \caption{Generic space $\gamma$-ray detector}
 \label{fig:generic}
\end{figure}

The complication is the fact that these charged particles interact with the converter and any other material in their paths through multiple Coulomb scattering, which degrades the information about their initial directions.  In a Gaussian approximation, for electrons and positrons the root-mean-square scattering angle $\theta_0$ in radians projected on a plane (so the full space angle would be $\sqrt{\mathrm{2}}\theta_0$) is given by  \cite{Lynch}

\begin{equation}
\theta_0=\frac{\mathrm{13.6 MeV}}{\beta\mathrm{cp}}\sqrt{\mathrm{x/X_0}}\left[{\mathrm{1 + 0.038\ln{(x/X_0)}}}\right]
\end{equation}

\noindent where $\beta\mathrm{c}$ and p are the velocity and momentum of the particle, x is the thickness of the scattering material in g cm$^{-2}$, and X$_0$ is the radiation length.  The scattering in a thick converter quickly erases most of the information about the incident photon arrival direction. The high-Z, short-radiation-length material that is advantageous for enabling pair production is now a disadvantage because it aggravates the multiple Coulomb scattering.  The solution to this problem is to break the converter into multiple thin layers, interleaved with particle tracking detectors.  A generic instrument concept is shown in Figure~\ref{fig:generic}. Various implementations of this basic idea have been used in space $\gamma$-ray detectors.

Even harder to measure than the arrival direction of the $\gamma$ rays is their polarization, although polarization would be a valuable astrophysical diagnostic for objects such as pulsars, supernova remnants, active galactic nuclei and $\gamma$-ray bursts.  For linear polarization, information about the plane of polarization is carried through pair production into the plane of the electron-positron pair \cite{Maximon_Olsen}.  The effect, however, is small, and like the direction of the pair itself, the plane of the pair is easily distorted by multiple Coulomb scattering.  Nevertheless, polarization has been detected by pair production at accelerators, e.g. \cite{Barbiellini}. See \cite{Bloser}, \cite{Buehler}, \cite{Bernard}, and \cite{Hunter} for discussions of this topic in the context of possible space $\gamma$-ray instruments. At present, it is fair to say that polarization is an unrealized potential of space high-energy $\gamma$-ray detectors. 

\section{Space Instrumentation Issues}

Applying the principles outlined in the previous section to a $\gamma$-ray detector in space involves a wide range of other considerations.  Many of these are logistical issues like mass, power, and money.  Others represent the challenges of the launch stresses and the space operating environment.  The most important design issue, however, is finding an efficient way to deal with the intense charged particle environment experienced by any space instrument.  

\subsection{Background Radiation in Space}

\begin{figure}[!t]
 \centering
 \includegraphics[width=15.5cm]{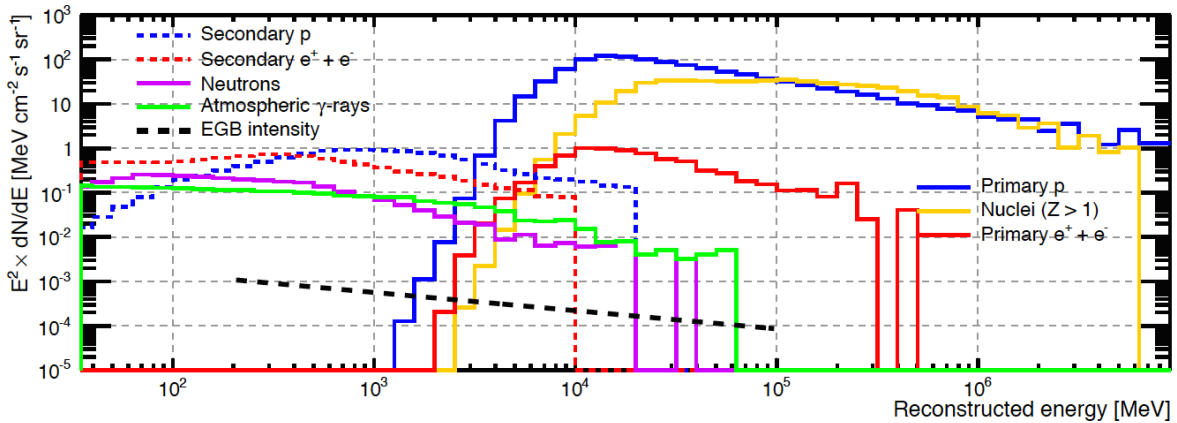}
 \caption{Model of the Fermi orbital position and particle direction-averaged
    radiation intensities \cite{2004.CR_MODEL} sampled from a
    background simulation run. The intensity of the
    extragalactic $\gamma$-ray background (EGB - the heavy dashed line in the figure) measured by the Fermi LAT
    \cite{ExtraGalBkg} is shown for comparison. The event energy
    is reconstructed under the hypothesis of a downward-going $\gamma$ ray
    and in general does not represent the actual energy for hadrons.}
 \label{fig:Background}
\end{figure}

Even within the protection offered by Earth's magnetic field, space is filled with intense radiation, including primary hadronic and leptonic cosmic rays, along with secondary protons, electrons,  neutrons, and $\gamma$ rays produced by cosmic-ray interactions in the atmosphere.  Figure~\ref{fig:Background} shows a typical background for a satellite in low-Earth orbit, based on modeling done for the Fermi Gamma-ray Space Telescope \cite{Pass7}.  The flux decrease for primary particles below 10 GeV reflects the geomagnetic cutoff.  Shown for comparison is the extragalactic $\gamma$-ray background, which is orders of magnitude less intense than the local background in this environment. Further, the background comes from all directions, since many of the secondary particles are coming up out of the atmosphere. Separating the cosmic $\gamma$ rays from all this unwanted radiation is the biggest problem in designing a successful space $\gamma$-ray instrument. 

The primary approach to identifying the charged particle background is to surround the active detector with an anticoincidence detector, as illustrated in Figure~\ref{fig:generic}.  Plastic scintillator is the material of choice, because it has high efficiency for detection of charged particles but low probability of absorbing $\gamma$ rays.  A signal in this detector implies a charged particle that can be rejected, either in hardware or software triggering of the instrument.  Note that the anticoincidence detector cannot surround the entire instrument, because at GeV energies the $e^+$ and  $e^-$ produce showers, including penetrating particles that would produce self-veto.  Even with an extremely efficient anticoincidence detector, however, there will be some particles that will go undetected or enter through uncovered parts of the geometry and must be removed by other means.  Space $\gamma$-ray instruments use every bit of information about each potential pair production event to determine whether it was a real cosmic photon interaction. The ability of an instrument to produce analyzable data, separating signal from background, is a critical design element.

\subsection{Reaching and Surviving the Space Environment}

In addition to achieving science goals, the design of any space instrument must consider the realities of rocket launches and operating in space.  Rockets generate noise, vibrations, and large sustained acceleration.  Repairing an instrument once it is in space is essentially impossible.  Similarly, expendable supplies such as fuel, gas or cryogen are not likely to be replaced.  Durability is therefore a vital element of any space project.  Particularly in low-Earth orbit, a spacecraft faces a challenging thermal situation.  Many times each day, the satellite makes a transition from being in full Sun to facing exposure to the cold of deep space.  Thermal stresses, as well as changes in performance due to temperature variations, must be considered. All this must be done in a vacuum, where radiative transport, rather than conduction or convection, is the principal form of heat transfer.  The space radiation environment requires that electronic circuits be radiation-hardened to minimize failures.  Such design considerations are common to any space instrument. 

As an example of the sort of trade-off required, consider natural micrometeoroids and human-generated space debris, a problem that has been exacerbated by some destructive collisions of satellites.  Micrometeoroid/space debris objects have a wide range of sizes, with potential impact velocities often reaching 20 km/s or more, compared to 1-2 km/s for a high-power rifle bullet. Such high velocities make even small projectiles dangerous to satellites\footnote{See http://esabase2.net/wp-content/uploads/2013/07/ESABASE2-Debris-Technical-Description.pdf}.  The outer element of a $\gamma$-ray detector, as noted above, is likely to be plastic scintillator, which must be shielded from light.  Putting a massive barrier outside the anticoincidence detector, however, negatively impacts the science, both by absorbing incoming photons and by acting as a $\gamma$-ray background source from inelastic collisions of cosmic-ray particles with the barrier material (producing neutral pions that decay immediately to $\gamma$ rays). The compromise is to use a low-Z multi-layer composite blanket that has a high probability of stopping projectiles of a size and velocity likely to hit the instrument during its lifetime.

\subsection{Practical Issues of Space Instrumentation}

No matter how excellent the scientific motivation, the ultimate existence of any satellite project depends on obtaining funding from agencies like NASA or ESA.  That hard fact means that the design must be optimized to keep the cost reasonable.  Managing the cost implies constraining the mass and power (because producing more power means more mass, as in larger solar panels).  The cost of putting each kilogram of  satellite mass into orbit is many thousand dollars or Euros.  Similarly, the overall size of an instrument is limited, both by the mass and fitting into a launcher (larger launchers are more expensive).  Data downlink bandwidth may also be limited (by power onboard and access to ground stations).    Reliability of long-term operation in space can only be assured by extensive (and expensive)  testing before launch, simulating launch and on-orbit conditions as accurately as possible.  In many cases, limits on some or all these parameters are dictated by the flight opportunities.  The challenge is always to maximize the scientific return within all these constraints. 

\section{Examples}
Three generations of space $\gamma$-ray instruments have been flown:
\begin{enumerate}
\item The $\gamma$-ray detector on the OSO-3 satellite was a sophisticated counter telescope. It provided the first definitive measurement of high-energy cosmic $\gamma$ radiation\cite{OSO-3}.  Although it did not image the pair production events, it nevertheless showed that the Galactic plane is a bright $\gamma$-ray source. 
\item Balloon payloads, then the SAS-2, COS-B, and CGRO/EGRET satellite instruments represented the second generation of instrumentation, using 1960s and 1970s technology to measure pair-production $\gamma$ rays.
\item AGILE and Fermi LAT are the current state of the art in high-energy $\gamma$-ray astrophysics, based on modern techniques used in particle physics.
\end{enumerate}

The following sections describe how the principal satellite instruments of the second and third generation dealt with the scientific and operational challenges described in previous sections. 

\subsection{EGRET and Earlier Instruments}

The first $\gamma$-ray instruments with particle-tracking capability were flown on balloons in the early 1970s, confirming the bright emission from the Galactic plane \cite{Fichtel_balloon} and finding evidence for pulsed $\gamma$ rays from the Crab pulsar, e.g. \cite{Vasseur}, \cite{Browning}, \cite{Albats}.  These were followed by two small satellites:  the Second Small Astronomy Satellite (SAS-2, 1972$-$73), which mapped the Galactic emission \cite{Fichtel1975}, discovered $\gamma$-ray emission from the Vela pulsar \cite{Thompson1975}, and measured an isotropic background radiation \cite{Fichtel1978}; and the COS-B satellite (1975$-$82), which produced a catalog of high-energy $\gamma$-ray sources, 2CG \cite{Swanenburg}, including 3C273, the first extragalactic source \cite{Bignami}, as well as much greater detail on the pulsars and Galactic $\gamma$ radiation, e.g. \cite{Kanbach},\cite{Hans}, \cite{Lebrun}.  Although the instrument configurations varied, all used spark chambers as the tracking detector. Spark chambers were rugged, reliable, low-power detectors that were easily adapted to space operation.

The success of these pioneering $\gamma$-ray detectors set the stage for a major project, the Compton Gamma Ray Observatory (CGRO), a NASA mission with international partnerships.   The four instruments - Burst and Transient Source Experiment (BATSE), Oriented Scintillation Spectrometer Experiment (OSSE), Compton Telescope (COMPTEL), and Energetic Gamma Ray Experiment Telescope (EGRET) - provided a comprehensive view of the $\gamma$-ray sky from 30 keV to greater than 10 GeV.  The high-energy instrument, EGRET, is the focus of the following. 

\begin{figure}[!t]
 \centering
 \includegraphics[width=9.5cm]{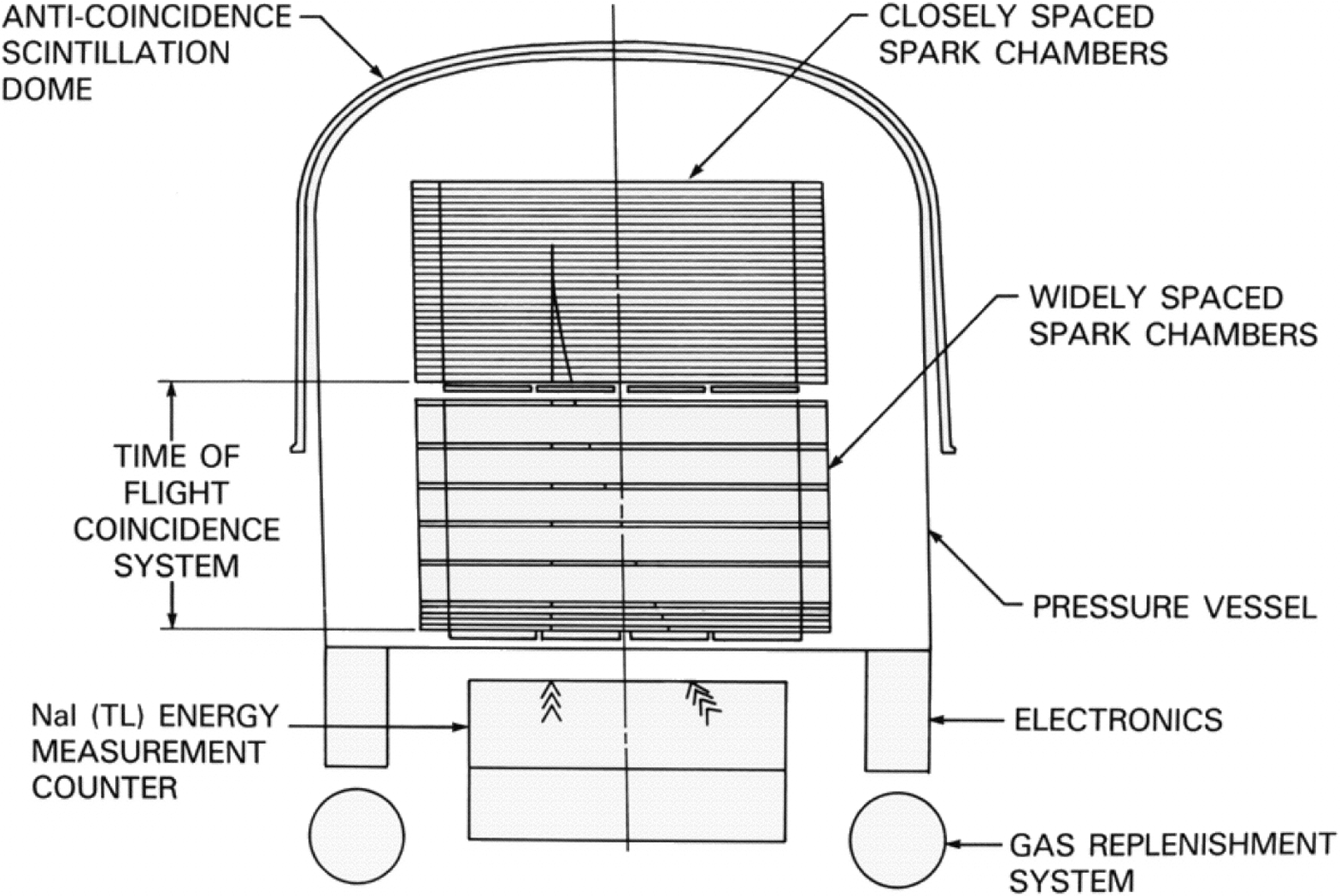}
 \includegraphics[width=6.0cm]{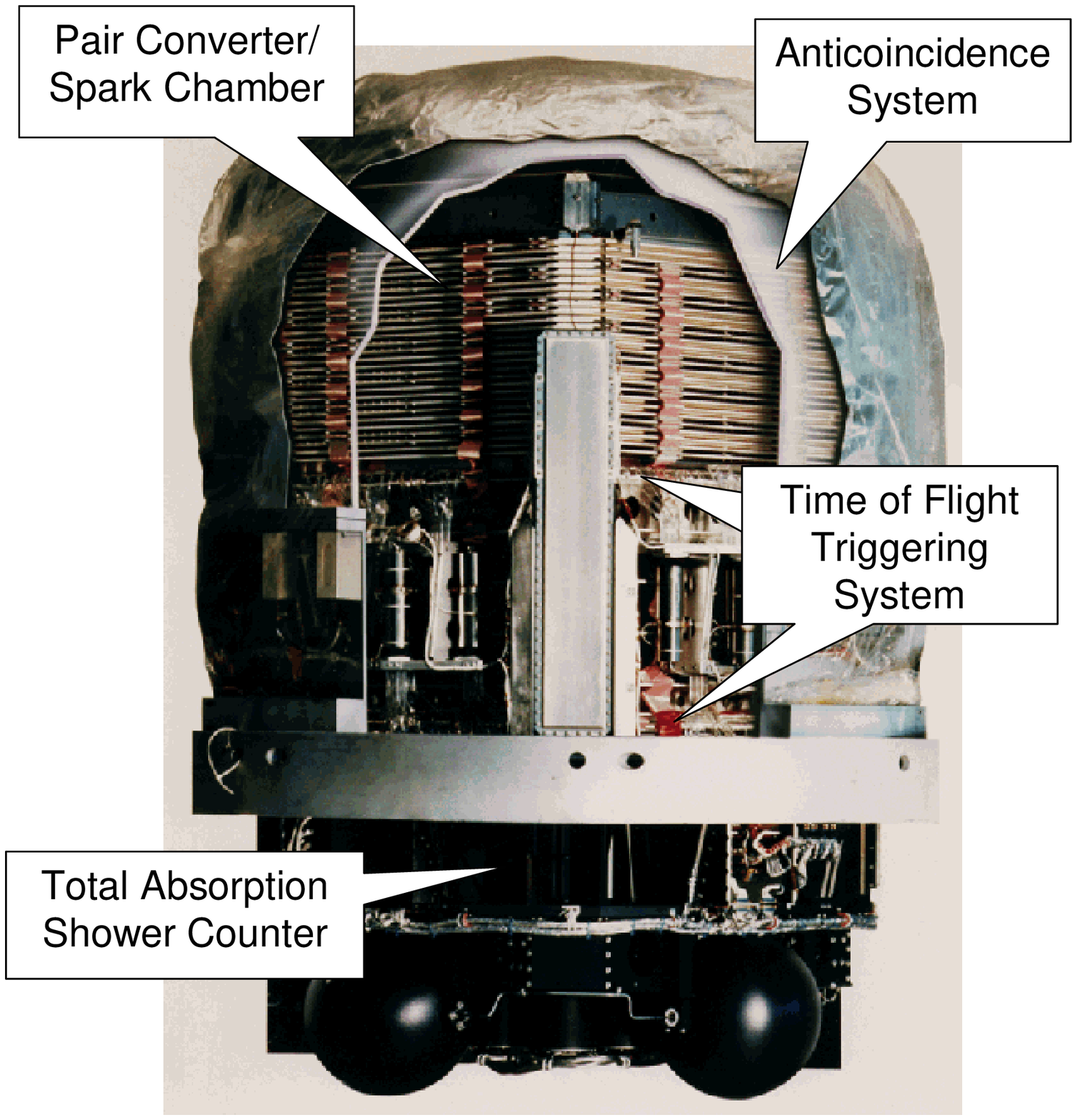}
 \caption{Left: Schematic diagram of EGRET \cite{Esposito}.  Right: composite photo of EGRET, showing the various components. }
 \label{fig:EGRET_schematic}
\end{figure}

In many respects, EGRET (Figure~\ref{fig:EGRET_schematic}) was the culmination of the experience gained from the balloon and small satellite projects, using many of the same technologies. In particular, spark chambers were used for the tracking detectors.  By 1978, when EGRET was proposed, the high-energy physics community had largely moved away from spark chambers in favor of drift chambers and multiwire proportional chambers, which offered higher spatial resolution.  For a space application, however, the low power of spark chambers became the deciding factor (the power allocated to each of the instruments on the observatory was less than 200 watts).  EGRET was a collaborative effort by NASA Goddard Space Flight Center, Stanford University, and the Max Planck Institute for Extraterrestrial Physics. 

The EGRET implementation of the basic principles described above involves several subsystems and structures \cite{Kanbach_EGRET}:

\begin{itemize}
\item The converter material was a stack of 27 80 cm x 80 cm tantalum foils of 0.02 radiation length thickness, providing a reasonable detection efficiency while minimizing the effects of multiple Coulomb scattering.
\item The tracking of the $e^+$ and  $e^-$ was done by a stack of wire grid spark chambers in a sealed vessel filled with spark chamber gas (neon/argon/ethane).  28 of these were interleaved with the conversion foils, with an additional 6 chambers below with no converters, to help identify and track the particle paths.  Each chamber had 992 wires in the x direction and 992 in the y direction, where z represents the vertical, pointing direction of the instrument. Each wire threaded an individual magnetic core. The charged particles left an ionization path in the gas.  When high voltage was applied to one wire plane and the other was grounded, a spark closed the circuit, with current flowing down an x wire, across the spark to a y wire, and back down the y wire, setting the cores.  Following a trigger, the time to read out the core array and recharge the high-voltage pulsers was 110 milliseconds. 
\item The triggering for the spark chamber was provided by two 4 x 4 arrays of plastic scintillator tiles, operated in coincidence with each other. The large plastic scintillator dome that surrounded the upper part of the instrument provided the anticoicidence.  The readout was photomultiplier tubes.  Because the response of plastic scintillator is fast, it was possible to use hardware logic to make a decision about the trigger in less than 1 microsecond.  In order to discriminate against unwanted upward-moving particles, a time-of-flight requirement was included, to select particles hitting the upper scintillator plane first.  The triggering tile array was also used to change the EGRET field of view dynamically, optimizing the exposure to the sky while avoiding the bright Earth limb.  
\item  The energy of the electron-positron pair was measured by a sodium iodide crystal array,  8 radiation lengths thick, bonded into a monolithic detector and read out by phototubes.  In addition to the primary energy range above about 20 MeV, the energy measurement system had an independent readout in the 0.6$-$140 MeV range, used for solar flare and $\gamma$-ray burst studies. 
\item The spark chambers, time-of-flight triggering system, and some of the electronics were all housed in a thin aluminum pressure vessel, filled with 1.1 atmospheres of spark chamber gas.  Because spark chamber gas gradually deteriorates in use, high-pressure tanks carried enough gas to replace the entire volume four times. 
\end{itemize}

Monte Carlo simulations of EGRET provided the initial performance parameters.  The full instrument was then calibrated at accelerator facilities \cite{Thompson_calibration}.  The primary calibration was done at the Stanford Linear Accelerator Center (SLAC) using a high-energy electron beam that scattered optical laser photons to $\gamma$-ray energies (Compton scattering).  Exposures to photon energies from 15 MeV to 10 GeV and incidence angles from 0$^{\circ}$ to 40$^{\circ}$ provided a matrix of calibration points covering essentially the entire range of operation for EGRET.  A proton calibration to test for background production/detection was done at Brookhaven National Laboratory, and a follow-up $\gamma$-ray calibration was carried out at the Bates Linear Accelerator.  

The experimental calibrations provided empirical input for the EGRET data analysis system.  Although the on-board triggering system was very efficient, the data sent to the ground still contained unwanted or unusable triggers.  An automated scan of the triggers rejected those with too little information or patterns inconsistent with a $\gamma$-ray pair production event, then attempted to structure the spark chamber pattern into tracks, deriving a direction and energy.  The results were reviewed by data analysts who had trained using the calibration data sets.  The basic criterion was to be consistent with a topology of a pair of tracks originating from a common point (the pair production itself), or if the energy were high enough, single straight tracks originating within (not at the edge of) the spark chamber volume.  Tests on flight data using known sources and comparison with the particle flux at different points in the CGRO orbit showed that the final data had no measurable contamination from non-$\gamma$-ray triggers \cite{Esposito}, \cite{Bertsch_calibration}, \cite{Sreekumar}.

\begin{figure}[!t]
 \CenterFloatBoxes
\begin{floatrow}
\ttabbox
{\begin{tabular}{@{}ll}
Property&Value\\
\hline
Energy Range&20 MeV $-$ $>$10 GeV\\
Peak Effective Area&1500 cm$^2$ at 500 MeV \\
Energy Resolution&15\% FWHM\\
Effective field of view&0.5 steradians\\
Timing accuracy&$<$ 100 $\mu$sec absolute\\
\end{tabular}
}
{\caption{\label{label}Some Performance Characteristics of EGRET}
}
\killfloatstyle
\ffigbox
 {\includegraphics [height = 3.5 in., angle =90] {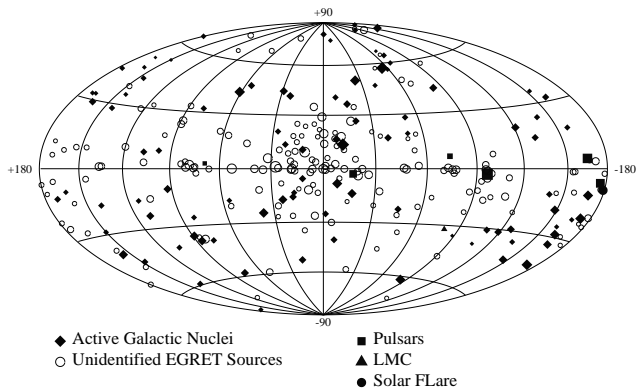}}
{\caption{Map of E $>$100 MeV sources in the third EGRET catalog, 3EG \cite{3EG}, shown in Galactic coordinates.  The symbol size indicates the source brightness. }
\label{fig:3EG}}
\end{floatrow}
\end{figure}

Table 1 summarizes some of the key EGRET operational parameters.  The single-photon angular resolution, or point spread function, was energy-dependent.  The point-spread function (PSF) angle $\theta_{67}$ in degrees containing 67\% of the $\gamma$ rays from a delta function source with energy E in MeV  was given by:  $\theta_{67}$ = 5.85 (E/100)$^{-0.534}$ \cite{Thompson_calibration}.

The Compton Observatory was launched on the Shuttle Atlantis in April 1991, with a planned 2-year mission.   It operated in pointing mode, with typical exposures of two to three weeks to allow adequate exposures. All the instruments were still operating when the mission finally ended in June, 2000, following the failure of a reaction wheel.  At that time, the massive (17000 kg) observatory was de-orbited into the Pacific Ocean as a safety measure. 

EGRET produced the first map of the entire high-energy $\gamma$-ray sky and demonstrated the diversity of energetic phenomena detectable at these energies.  Sources ranged from solar flares just 8 light-minutes from Earth to the diffuse emission of the Milky Way to $\gamma$-ray bursts at cosmological distances.  See \cite{Thompson_review} for an overview of EGRET results.  Much of the EGRET scientific legacy is captured in the Third EGRET Catalog (3EG, \cite{3EG}), shown in Figure~\ref{fig:3EG}.  Important source classes were active galactic nuclei (mostly of the blazar class) and pulsars, with hints of other classes such as supernova remnants and binary systems containing a compact object.  Over half the 271 sources had no association with plausible $\gamma$-ray-producing astrophysical objects. Some of these were likely gas clouds that were not included in the diffuse model used for the analysis \cite{Grenier}.  Others were later found to have likely associations with known objects, e.g. \cite{3EG_blazars}.

\subsection{Fermi LAT}

By the time the Compton Observatory had completed its first all-sky survey, the early  success of that mission had led the Stanford/SLAC group to start planning a follow-on high-energy mission, taking advantage of the advances in detector technology and modeling in the years since EGRET was designed \cite{Atwood}.  That work led to the Large Area Telescope (LAT) \cite{Atwood_LAT} on the Gamma-ray Large Area Space Telescope (GLAST), renamed after launch in 2008 to Fermi Gamma-ray Space Telescope.  Fermi LAT represents an international/interagency collaboration, including NASA, the Department of Energy, and universities in the U.S. along with universities and scientific agencies in France, Italy, Sweden, and Japan.  A smaller Italian mission, Astro-rivelatore Gamma a Immagini LEggero (AGILE), uses similar technology to the LAT \cite{AGILE}.

The single most important development between the EGRET and Fermi-LAT eras was the emergence of large-area silicon strip tracking detectors.  Coupled with extremely low-power Application-Specific Integrated Circuits (ASICs), silicon strip detectors offered significant advantages over the older gas detectors:  higher spatial resolution, no pressure vessel or gas degradation, self-triggering, much faster readout, and extremely high efficiency.   

Almost as significant was the dramatic increase in computing power.  Simulations at the individual detector readout level became possible, allowing not only optimization of the instrument design but also parallel development of data analysis software to reject background and extract information from the pair production particles.  Some onboard analysis could be done, giving far more flexibility in data acquisition than the largely hardware-dominated EGRET system. 

\begin{figure}[!t]
 \CenterFloatBoxes
\begin{floatrow}
\ffigbox
 {\includegraphics[width=7.5cm]{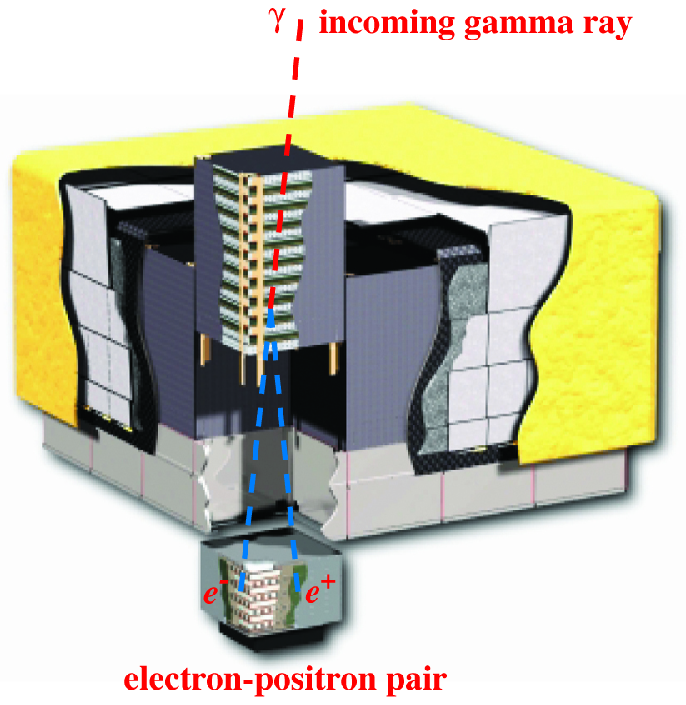}}
 {\caption{ Schematic diagram of Fermi LAT \cite{Atwood_LAT}.}   }
{ \label{fig:LAT}}
\killfloatstyle
\ttabbox
{\begin{tabular}{@{}ll}
Property&Value\\
\hline
Energy Range&20 MeV $-$ $>$300 GeV\\
Peak Effective Area&8000 cm$^2$ above 10000 MeV \\
Energy Resolution&$<$15\% FWHM\\
Effective field of view&2.4 steradians\\
Timing accuracy&$<$ 1 $\mu$sec absolute\\
\end{tabular}
}
{\caption{\label{label}Some Performance Characteristics of Fermi LAT}}
\end{floatrow}
\end{figure}

The Fermi LAT (Figure~\ref{fig:LAT}) implementation of the basic principles differs somewhat from the EGRET design, taking advantage of the new technologies available \cite{Atwood_LAT}:

\begin{itemize}
\item The tungsten converter foils are divided into two sections.  The 12 upper (front) layers are each 0.03 radiation lengths thick (to minimize Coulomb scattering effects that are important at lower energies), while four lower (back) layers have 0.18 radiation length thickness (to provide more converter for high-energy $\gamma$ rays that are less affected by scattering). 
\item The trackers are 18 x-y pairs of single-sided silicon strip detectors with strip pitch of 228 $\mu$m, with the first 16 interleaved with the converter foils.  The trackers are modular, with 16 towers in a 4 x 4 array; the total number of data channels is approximately 880,000.  The tracker system self-triggers on a pattern of three consecutive x-y pairs producing a signal. Not requiring a separate time-of-flight coincidence trigger allows the LAT to be more compact, increasing its field of view, although the absence of a time-of-flight signal complicates the background rejection. 
\item  The energy measurement is carried out by a calorimeter of 1536 cesium iodide logs, stacked in 8 layers with directions alternating x and y, read out by custom photodiodes. The calorimeter modules are arranged in the same 4 x 4 array of towers  as the tracker.  The hodoscopic structure allows the shape of high-energy showers to be measured, enabling measurements to higher energies than was possible with the EGRET system. 
\item  Unlike the single-piece EGRET anticoincidence scintillator, the LAT system consists of 89 individual scintillator tiles, overlapped, with scintillating fibers covering seams.  EGRET was subject to self-veto, especially at high energies, when backward-moving secondary particles from electromagnetic showers hit the scintillator.  With the LAT, only tiles close to the direction of the incoming $\gamma$ ray need to be examined.
\item Although the EGRET hardware trigger was a bit faster than that of the LAT, the LAT readout is dramatically faster, with only 26 $\mu$s dead time.  This improvement permits LAT to handle much higher trigger rates.  Coupled with the much larger telemetry bandwidth available to the LAT, this high rate enables much more information to be transmitted to the ground. 
\end{itemize}

The design and performance parameters for the Fermi LAT were optimized with extensive Monte Carlo simulations.   Accelerator beam tests were carried out on an engineering model \cite{Atwood2000} and a calibration unit consisting of a subset of flight-design detectors \cite{Baldini}.  The same type of simulation package used to design the instrument was used to develop automated analysis procedures for the data, eliminating the manual review that had been needed for EGRET.  The automated analysis separated the pair production events from unwanted triggers and determined the arrival direction and energy for each event.  Having an automated system has also allowed the LAT team to re-analyze data as improved knowledge of the instrument and its space environment have become available, e.g. \cite{Pass7},\cite{Pass8}.  Table 2 shows some of the key performance parameters of the Fermi LAT.   An approximation for the PSF 68\% containment angle  $\theta_{68}$ in  degrees for $\gamma$ rays with energy E in MeV converting in the front (thin-converter) section of the LAT is given by \cite{LAT_PSF}

\begin{equation}
\theta_{68}=\sqrt{(\mathrm{3.5}(E/100)^{-0.8})^2 + \mathrm{0.15}^2}
\end{equation}

\noindent Back-converting $\gamma$ rays have a PSF angle about twice as large.  At low energies, the LAT PSF is similar to that of EGRET, but the LAT PSF improves faster with increasing energy.  Current details about LAT performance parameters can be found at \url{http://www.slac.stanford.edu/exp/glast/groups/canda/lat_Performance.htm}.

The Fermi observatory has spent most of its observing time in scanning mode, with the LAT looking away from Earth.  Taking advantage of the huge LAT field of view, the typical pattern is to rock the satellite south of the orbital plane for one 96-minute orbit, then rock north for the next orbit.  In this way, LAT achieves exposure to the entire sky every 2 orbits. During passages through the high-particle-intensity South Atlantic Anomaly, triggering is disabled. 

\begin{figure}[!t]
 \centering
 \includegraphics[width=14.5cm]{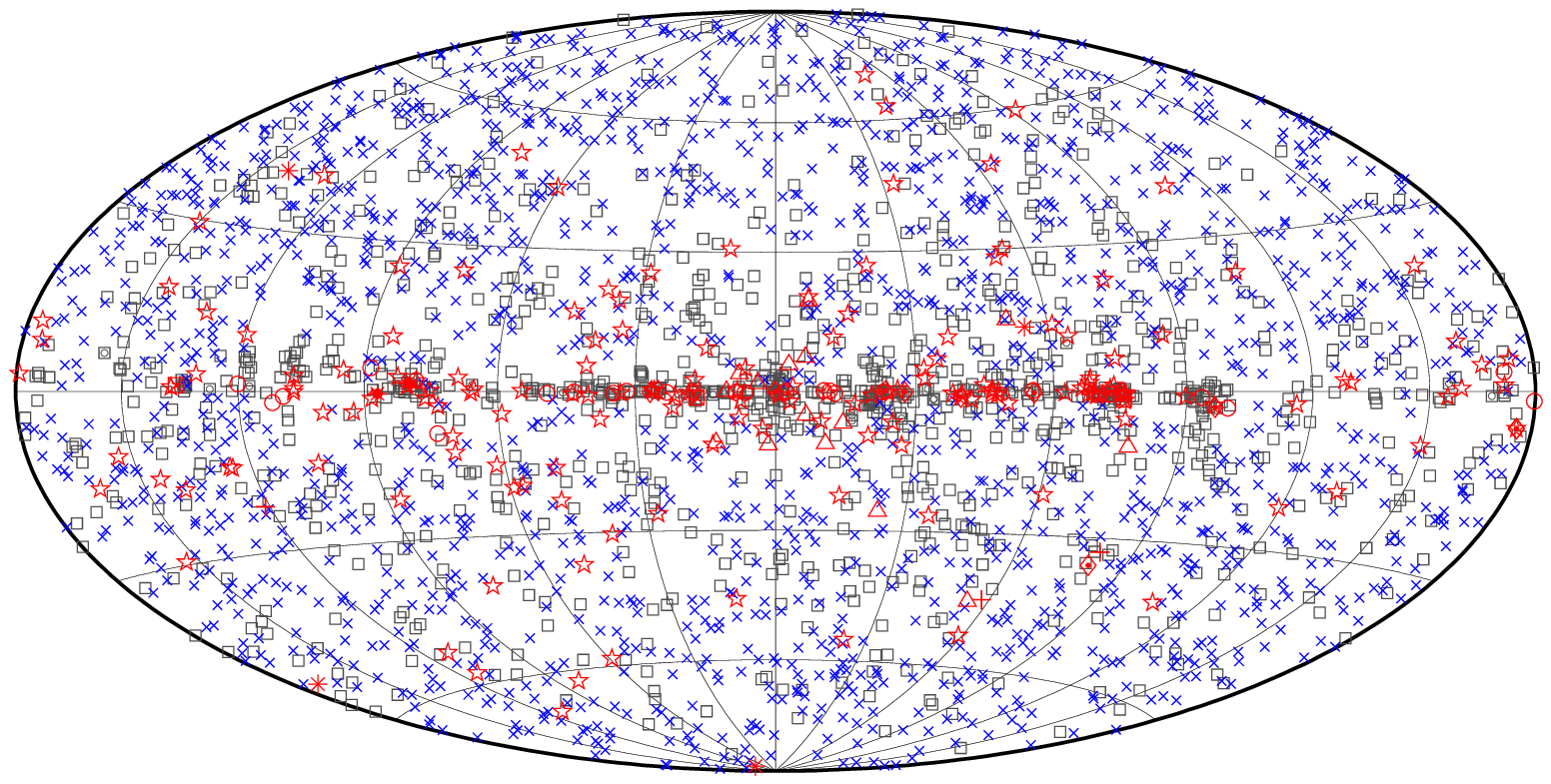}
 \includegraphics[width=14.5cm]{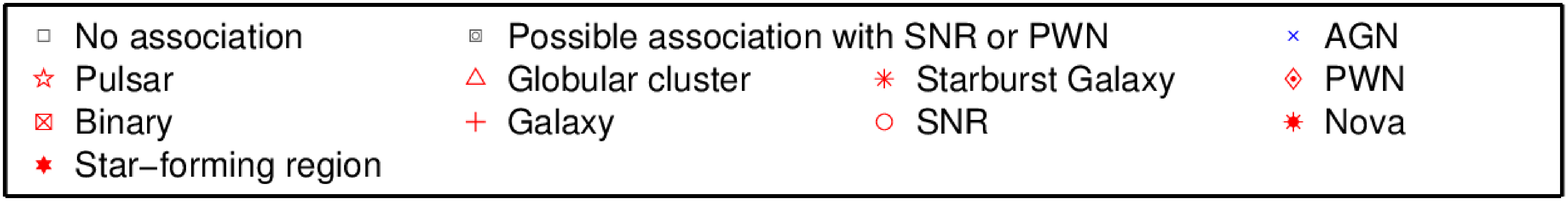}
 \caption{Fermi LAT third catalog \cite{3FGL}. }
 \label{fig:3fgl}
\end{figure}

The combination of large effective area, huge field of view, and scanning mode of operation has allowed LAT to produce a $\gamma$-ray data set vastly larger than all previous satellite missions combined.  The Third Fermi LAT Catalog (3FGL, \cite{3FGL}), shown in Figure~\ref{fig:3fgl}, has more than an order of magnitude more sources than the 3EG catalog.  Discoveries have included large-scale features like the Fermi bubbles \cite{Bubbles}, new source classes (e.g. $\gamma$-ray novae \cite{novae}) and an unprecedented level of detail about known sources.  AGILE has also contributed to the growing field of high-energy $\gamma$-ray  astrophysics.  Many of the other entries in this volume discuss  the results from these satellite instruments.

\section{Conclusion}

Satellite instrumentation for E$>$100 MeV $\gamma$-ray  astrophysics has advanced tremendously since the first detections by OSO-3.  The field now covers a broad range of topics, ranging from local phenomena in our Solar System to objects like blazars and $\gamma$-ray bursts at cosmological distances.  At the time of writing, Fermi and AGILE are both still operating well and are capable of many more years of useful observations.  There are no major technological developments on the horizon that seem capable of making the same sorts of major advances achieved in past missions.  Future satellite $\gamma$-ray projects appear likely to focus on specialized areas such as lower energies, higher angular resolution, and sensitivity to polarization.  
The current generation of instruments has by no means exhausted their capabilities.  New discoveries in this energy range are still expected. 


\section*{Acknowledgements}
Space missions like the ones described here involve many scientists, engineers, technicians, managers, and others.  I extend thanks to all those who have made space $\gamma$-ray astrophysics the dynamic field that it is today.  Special thanks go to Robert Hartman for valuable suggestions during the preparation of this article. 

Figures 3, 4, 5, 6, and 7 are reproduced by permission of the American Astronomical Society.


\end{document}